\documentclass{LMCS}

\def\doi{8 (1:30) 2012}
\lmcsheading%
{\doi}
{1--26}
{}
{}
{May~\phantom.19, 2011}
{Mar.~27, 2012}
{}

\usepackage{amssymb}
\usepackage{alltt}
\usepackage{graphicx}
\usepackage{url}
\usepackage{enumerate,hyperref}

\let\xmedskip=\medskip
\def\xtoolong{\hspace{-100cm}}
\def\toolong{$\xtoolong$}
\def\To{\Rightarrow}
\def\tskip{{\hspace{1pt}}}
\def\tstrut{{\vrule height8.5pt depth3pt width0pt}}
\def\tbar{\hspace{.5pt}\vrule height9.5pt depth4pt width.3pt\hspace{.5pt}}
\def\lam{\char`\\}
\def\{{\char`\{}
\def\}{\char`\}}
\def\treturn{$\hspace{.5pt}\langle{\it return}\rangle\hspace{-1pt}$}
\def\treturn{\hspace{-.5pt}}
\def\creturn{$\hspace{.5pt}\langle\mbox{\tt \char`\^C}\;{\it return}\rangle\hspace{-.5pt}$}
\def\creturn{$\mbox{{\it control-}\tt C}\,{\it return}$}

\def\notion#1{\,\langle\mbox{\it #1}\rangle\,}
\def\terminal#1{\,\langle\mbox{\rm #1}\rangle\,}
\def\key#1{\,\fbox{\strut\tt #1}\,}
\def\literal#1{\,\fbox{\strut\tt #1}\,}
\def\alter{\phantom{::=}\llap{$|$}}

\begin{document}

\title[A Synthesis of the Procedural and Declarative Styles of Interactive \dots]
{A Synthesis of the Procedural and Declarative Styles of
Interactive Theorem Proving}
\author[F.~Wiedijk]{Freek Wiedijk}
\address{
Institute for Computing and Information Sciences,
Radboud University Nijmegen,
Heyendaalse\-weg~135, 6525 AJ Nijmegen, The Netherlands}
\email{\texttt{freek@cs.ru.nl}}

\begin{abstract}
We propose a synthesis of the two proof styles of interactive
theorem proving: the procedural style (where proofs are
scripts of commands, like in Coq) and the declarative style
(where proofs are texts in a controlled natural language,
like in Isabelle/Isar).  Our approach combines the advantages
of the declarative style -- the possibility to write formal
proofs like normal mathematical text -- and the procedural
style -- strong automation and help with shaping the proofs,
including determining the statements of intermediate steps.

Our approach is new, and differs significantly from the ways
in which the procedural and declarative proof styles have
been combined before in the Isabelle, Ssreflect and Matita systems.
Our approach is generic and can be implemented on top of
any procedural interactive theorem prover, regardless of
its architecture and logical foundations.

To show the viability of our proposed approach, we fully
implemented it as a proof interface called \texttt{miz3}, on top of
the HOL Light interactive theorem prover.  The declarative
language that this interface uses is a slight variant
of the language of the Mizar system, and can be used for
any interactive theorem prover regardless of its logical
foundations.  The \texttt{miz3} interface allows easy access to
the full set of tactics and formal libraries of HOL Light,
and as such has `industrial strength'.

Our approach gives a way to automatically convert any
procedural proof to a declarative counterpart, where
the converted proof is similar in size to the original.
As all declarative systems have essentially the same proof
language, this gives a straightforward way to port proofs
between interactive theorem provers.
\end{abstract}

\maketitle

\section{Introduction}

\subsection{Proof styles of interactive theorem proving}\label{problem}

\noindent
Interactive theorem provers, also known as proof assistants,
are computer programs for the development and verification
of mathematical texts in a formal language.
These systems make
it \emph{certain}%
\footnote{
There is only one {serious} possibility for mathematics verified with the best interactive theorem provers to still
have problems \cite{pol:98,hal:11}.
The {definitions} and statements might not {mean} what the
person who wrote them thinks they mean.
Although it then still is \emph{certain} that the mathematics contains no errors at all, it is the `wrong'
mathematics.
}
that the verified mathematics contains no errors at all.
The activity of coding mathematics in the formal language of an interactive
theorem prover is called \emph{formalizing}, and the set of resulting input
files for such a system is called a \emph{formalization}.
Highly non-trivial proofs have been formalized, both in mathematics \cite{gon:06,har:08}
and in computer science \cite{fox:03,kle:09,ler:06}.

In interactive theorem proving one can consider the proofs on two different levels.
There is the \emph{user level proof}, the proof in the formalization files
on the level of which the user
interacts with the system.
And there is the \emph{proof object}, the proof in the formal system
underlying the system.
Generally the second is an order of a magnitude larger than the former.
In systems like Coq and HOL \cite{gor:mel:93}, the user level proof consists of
a list of tactics to be executed.
The proof objects in Coq are lambda terms, while in HOL they
consist of traces of function calls into the LCF style kernel of the system.
In a system like Mizar \cite{gra:kor:nau:10} the user level proof consists of
the proof steps in the input language of the system.
The Mizar implementation does not keep track of proof objects, but the proofs
on the proof object level
would be the formal deductions in first order predicate logic.

Interactive theorem provers can use three different proof styles
(the terminology originates in \cite{har:96:2}):
\begin{enumerate}[(1)]
\item
\emph{The procedural style.}
\label{item:procedural}
In these systems the user inputs a proof as a sequence of \emph{tactic} invocations, which
are commands that transform proof obligations called \emph{goals}.
A tactic reduces a goal to zero or more new subgoals.
When all goals have been solved this way, the proof is finished.
Note that although most procedural systems support both
forward and backward proof,
the {user interaction} in those system primarily consists of reasoning backwards from a goal.
Systems that use the procedural style are
the various HOL systems like HOL4 \cite{gor:mel:93}, HOL Light \cite{har:xx,har:00} and ProofPower \cite{lem:00}, the original version of Isabelle \cite{nip:pau:wen:02}, Coq \cite{coq:10}, Matita \cite{asp:coe:tas:zac:07}, PVS \cite{owr:rus:sha:92}, the B method \cite{abr:96} and Metamath \cite{meg:97}.

Some procedural systems offer the option to print their proof
objects in the form of natural language text.
An example of such a system is Matita (see the discussion in Section~\ref{related} below).
\smallskip

\item
\emph{The declarative style.}
\label{item:declarative}
In these systems the user inputs proofs in a stylized natural deduction language
(de Bruijn called such a language a \emph{mathematical vernacular} \cite{bru:87}).
The output of the system then consists of messages that point
out where the proof text still has errors.
There are two subclasses of this proof style:
\smallskip

\begin{enumerate}[(a)]
\item
\label{item:declarative:natlang}
\emph{The natural language declarative style.}
Here the proof language is a formal version of mathematical natural language,
also called a \emph{controlled} natural language.
Trivial reasoning between the steps in a proof is provided by the system through
(light weight) automation.
Systems that use this proof style are Mizar \cite{gra:kor:nau:10} and Isabelle with its `structured' proof language Isar \cite{wen:02,wen:02:1}.

Actually, the Mizar and Isar languages, as well as the language described in this paper, are not very much like natural language.
The ForTheL language for the SAD system \cite{pas:07} is much better in this respect.
\smallskip

\item
\emph{The proof object declarative style.}
\label{item:declarative:object}
Here the proof input language is a syntactic rendering of the proof object,
with the structure of a natural deduction proof.
Systems that use this style are
Twelf \cite{pfe:sch:02}, Agda \cite{agd:xx} and Epigram \cite{mcb:mck:04}.

In some of these systems, the user does not need to type the whole
input text by themselves, but can also give \emph{commands} in the interface that generate part of the proof text.
These commands will \emph{not} be part of the proof files that
are the final formalization.
Examples of such systems are Agda and Epigram.

\end{enumerate}
\smallskip

\item
\emph{The guided automated style.}
\label{item:guidedautomated}
In these systems the input is a sequence of lemma statements,
which the system then tries to prove by itself.
These systems generally produce long
natural language texts for each lemma describing how it was proved.
Often for some of the lemmas parameters need to be given that
direct the system how to perform the proof.
Also the lemma statements need to be chosen well for the system
to be able to do the proof, as the gaps between the statements should not
be too large.
For this reason these systems still
are \emph{interactive} theorem provers.
Systems that use this style are ACL2 \cite{kau:man:moo:00} and Theorema \cite{buc:jeb:kri:mar:vas:97}.

One can consider the guided automated style to be either an extreme version
of the procedural or an extreme version of the declarative style.
In a guided automated theorem prover, one runs one supertactic per
lemma.
Or, in a guided automated theorem prover one
writes a theory as a series of lemma statements, where the
system checks that each statement follows from the previous ones.

\end{enumerate}

\noindent
It is interesting to compare where the `proof commands to the system' and where the
`natural language proofs' are in these various proof styles:

\xmedskip
\begin{center}
\begin{tabular}{lcc}
& \emph{commands} & \emph{natural language} \\
\noalign{\smallskip}
\hline
\noalign{\smallskip}
procedural & input & absent or output \\
declarative, natural language & absent & input \\
declarative, proof object & absent or interface & absent \\
guided automated & input & output \\
\noalign{\smallskip}
\hline
\noalign{\smallskip}
{the system from this paper}
& interface \emph{and} input & input \emph{and} output \\
\noalign{\smallskip}
\hline
\end{tabular}
\end{center}
\xmedskip

\noindent
There are some systems that are
outside the simplicity of this
table, like Isabelle with its combination of natural language declarative
and procedural proof styles,
like Matita
with its declarative
proof checker, and like the `skeletonizer' of Mizar.
These systems will be discussed in Section~\ref{related} below.

The proof style that we propose in this paper 
is an integration of the procedural and natural language declarative
styles.
The advantage of the declarative style is that it is
closer to normal mathematical practice (one just writes proofs)
with more readable proof scripts,
and that it gives full control over the exact statements in the
proof.
Also declarative proofs tend to be easier to maintain
and less dependent on the specific system
than procedural ones.
The advantage of the procedural style is 
that one does not need to write all intermediate statements: these are generated automatically.
Also, procedural systems tend to have much stronger automation,
with often many different \emph{decision procedures}
that without human help can perform proofs in specific domains.

The goal of this paper is to propose a proof style that
combines the best of these two worlds.

\subsection{Relating the procedural and declarative proof styles}

\noindent
Looked at superficially, the procedural and declarative
proof styles seem very different.
Certainly the proof scripts for those two styles \emph{look} completely
different.
However, when working with these systems, both styles turn out
to have a very similar work flow.

When working on a declarative proof, most of the time when one
is not finished the only errors left are that
the system
did not succeed in proving some of the steps in the proof from the earlier steps.
In Mizar these steps are called \emph{unjustified} steps, and have error
numbers \texttt{*1} and \texttt{*4}.
Now these unjustified steps correspond exactly
to the subgoals that one looks at when a procedural proof is
not finished!
In other words, an unfinished proof of a Mizar lemma in which there still are -- say --
seven unjustified steps left, is very similar to a Coq or HOL
proof in which there are still seven subgoals left.
This is the first observation that is the basis of our approach.

A proof in a declarative system consists of `steps', most of which contain
a statement.
If one does the analogous proof in a procedural system, one goes
through many subgoals that \emph{also} consist of many statements.
Now it turns out that those two sets of statements in practice are very much the same!
In other words, if for a procedural proof we collect all the statements in the goals (both the assumptions and the statements to be proved),
then those statements can in a natural way be organized as a declarative
proof.
This is the second observation at the basis of our approach.

Note that in these two observations there is no reference to
proof objects.
This means that our integration of the procedural and declarative
proof styles does not have anything to do with proofs on the proof object level.
It also means that our proposed approach is independent of the
foundations or architecture of the system.
What we propose will apply to \emph{any} system in which the user
performs proofs by executing tactics on subgoals containing
statements.

Our proposal then is to have a proof interface in which a user is
working on a declarative proof.
In this proof the unjustified statements \emph{are} considered the subgoals of
the prover.
At any of these steps/subgoals one can execute any tactic of the system,
and if this is successful the statements of the new
subgoals that the tactic produces will be merged into the proof text,
making the declarative proof `grow' \cite{kal:wie:09}.
However, one also can freely manually edit and then recheck the declarative text.
The text does not need to `remember' how it has been grown.

For an example of how all this works out in a concrete session,
see Section~\ref{session} below.

\subsection{Related Work}\label{related}

\noindent
From the introduction in Section~\ref{problem} it will be clear
that the proof style that we propose is a combination of aspects of
many different proof systems.
However, there are some systems that are quite close
to what we propose.
For each of them we will discuss now how they differ from
our work:
\begin{desCription}
\item\noindent{\hskip-12 pt\bf Isabelle/Isar}:\ In Isabelle one can
  encapsulate procedural proof fragments consisting of tactic
  applications in a declarative Isar proof text
  \cite{wen:02,wen:02:1}.  However, the user needs to manually type
  the declarative text (it is not generated like in our approach), and
  the procedural proofs do not make sense without running them on the
  system (unlike in our approach, where the tactics do not
  \emph{solve} the goal but connect statements together).
\medskip

\item\noindent{\hskip-12 pt\bf Ssreflect}:\ The usage of Georges
  Gonthier's Ssreflect language for Coq \cite{gon:mah:tas:08} is
  similar to a common way of using Isar.  It is used declaratively for
  the high level structure of the proof while at the `leaves' of the
  proof the user switches to the procedural proof style.  However, the
  declarative part of Ssreflect is much less developed than Isar.
  Also, although Ssreflect is clearly intended to be also used
  declaratively, it barely fits category (2a) in the classification
  above.
\medskip

\item\noindent{\hskip-12 pt\bf HELM/MoWGLI/Matita}:\ The HELM, MoWGLI
  and Matita systems \cite{asp:coe:tas:zac:07,asp:03,asp:weg:02} have
  as one of their goals to render type theoretical proof objects as
  natural language.  In Matita, these rendered proof objects also can
  be read back in, and checked for correctness like in a declarative
  proof system \cite{coe:10}.

An important difference with our approach is that one cannot go
back from declarative editing to procedural proving.
Once a declarative proof text has been modified, if the
procedural proof from which it has been generated also gets modified,
both modifications cannot be integrated.
In other words, once one has worked declaratively, working procedurally
is no longer possible.

Another difference is that the declarative proof text is generated
from the proof object, which is generally more fine grained than
the user level proof on the level of the tactic invocations,
and is therefore more verbose and less understandable.
\medskip

\item\noindent{\hskip-12 pt\bf The proof rendering from the Lemme project}:\
When proofs are being rendered as natural language,
the source of the rendering is generally the proof object.
An important exception is a system
by Fr\'ed\'erique Guilhot, Hanane Naciri and Lo\"\i c Pottier.
Unfortunately, this work seems not to have been published,
all that exists is a set of slides for a talk about it \cite{gui:03}.

A difference with our approach is that the generated text cannot
be modified by the user anymore.
The rendering in this system is just output, and is not parsed back again.
\medskip

\item\noindent{\hskip-12 pt\bf NuPRL}:\
The NuPRL system \cite{con:all:bro:cle:cre:har:how:kno:men:pan:sas:smi:86} has a way to display formal proofs in which
groups of tactics are interleaved with fragments of goals.
Between the groups of tactics, the parts of the goal that have changed
are shown
(see for an example the NuPRL chapter in \cite{wie:06}).
This is quite similar to what happens in our approach.

But again, this rendering is just output, and is not parsed back again.
\medskip

\item\noindent{\hskip-12 pt\bf Mizar's skeletonizer}:\
In natural language declarative systems like Isar and Mizar, the proof
text has to be written by the user.
A slight exception to this is the `skeletonizer' of the emacs interface
to Mizar by Josef Urban \cite{urb:06:1}.
In this interface, a proof skeleton
is automatically generated from
the statement to be proved.

This is similar to what happens when we `grow' a proof by executing
a tactic.
However, in our case the growing is \emph{generic}: for each tactic
there is a corresponding way to insert part of the proof.
In Mizar there is only one such way.

\end{desCription}

\noindent
There already are various declarative proof languages which have been grafted
on top of a procedural system.
Currently Isabelle/Isar
is the only one that knows widespread use.
Others are:
\begin{desCription}
\item\noindent{\hskip-12 pt\bf `Mizar modes' for HOL Light}:\
There are two by John Harrison \cite{har:96,har:07:1} and two earlier ones by the author (\cite{wie:01} and an unpublished
one included in the HOL Light distribution \cite{har:xx}).
The \texttt{3} in the system name \texttt{miz3} refers to the fact that this is the third Mizar mode for HOL Light that we developed.
\medskip

\item\noindent{\hskip-12 pt\bf C-zar}:\ 
A declarative proof language for Coq by Pierre Corbineau \cite{cor:07}.
\medskip

\item\noindent{\hskip-12 pt\bf PhoX}:\
There exists an experimental declarative version of the PhoX theorem prover by Christophe Raffalli \cite{pho:xx}.

\end{desCription}

\noindent
These systems are all quite similar.
The main improvement of the work described in this paper over these
other systems is that it adds a Mizar-style interaction model,
and that it integrates execution of tactics with generation of proof text.

\subsection{Contribution}

\noindent
This paper is a continuation of the work in \cite{kal:wie:09,wie:01}.
It contains three contributions:
\begin{iteMize}{$\bullet$}
\item
We describe a declarative proof interface for the HOL Light
theorem prover that is much more developed and far more ergonomic
than earlier attempts at this.
This software can be downloaded at:
\xmedskip
\begin{center}
\url{http://www.cs.ru.nl/~freek/miz3/miz3.tar.gz}
\end{center}
\xmedskip

\item
We describe a new proof style for interactive theorem provers
that is a synthesis between the
procedural and declarative proof styles.

\item
We describe a method for automatically converting \emph{any}
existing procedural proof to a declarative equivalent.
This gives an approach for conserving
libraries of formal proofs and semi-automatically porting them between
systems.
For details see Section~\ref{automatic} below.

\end{iteMize}

\subsection{Outline}

\noindent
The structure of the paper is as follows.
In Section~\ref{session} we describe through an example how the proof interface that
we developed works.
In Section~\ref{language} we describe the declarative proof language of this interface.
In Section~\ref{implementation} we give some details of the implementation of this
interface.
In Section~\ref{automatic} we describe how our approach makes it possible
to automatically convert existing proofs to our language.
In Section~\ref{lagrange} we describe our experiences with using our interface
on a non-trivial example.
Finally in Section~\ref{discussion} we conclude with some observations and
planned future work.

\section{The \texttt{miz3} proof interface to HOL Light}\label{session}

\noindent
We developed a prototype of the interface style proposed
in this paper as a layer called
\texttt{miz3} on top of the HOL Light system \cite{har:xx,har:00}.
It consists of about 2,000 lines of OCaml code (in comparison,
the basic HOL Light system is approximately 30,000 lines), and its
development took 
three man months.

We will explain the \texttt{miz3} interface with a simple
example.
For this we will use the traditional inductive proof of
$$\sum_{i = 1}^n i = \frac{n (n + 1)}{2}$$
In HOL Light this is written as:
\xmedskip
\begin{center}\small
\texttt{!n.\ nsum (1..n) ({\lam}i.\ i) = (n * (n + 1)) DIV 2}
\end{center}
\xmedskip
In this formula, the exclamation mark \texttt{!} is ASCII for the universal
quantifier $\forall$, the backslash \texttt{\lam} is ASCII for the
$\lambda$ of function abstraction,
and the function \texttt{nsum} is a higher order version of the summation operator $\sum$.

Of course the example is very trivial.
We do not want to give the impression that our approach only works well
for simple examples like this.
We just chose this example to be able to fit a reasonable representation
of a proof session for it in the paper.
We worked on much larger proofs with \texttt{miz3},
and it performs well on those too.
For a description of such a larger example see Section~\ref{lagrange} below.

To prove the equality of the example, we will need one lemma,
which is the recursive characterization of \texttt{nsum}.
The statement of this lemma, which we will use for rewriting expressions
involving \texttt{nsum}, is:
\xmedskip
\begin{alltt}\small
# \fbox{NSUM_CLAUSES_NUMSEG;;\treturn\tstrut}\smallskip
val it : thm =
  |- (!m. nsum (m..0) f = (if m = 0 then f 0 else 0)) /{\lam}
     (!m n.
          nsum (m..SUC n) f =
          (if m <= SUC n then nsum (m..n) f + f (SUC n) else nsum (m..n) f))\toolong
\end{alltt}
\xmedskip
This is the first example of a command in a HOL Light session.
We indicate user input by putting boxes around it, to differentiate
it from the output from the system that is outside those boxes.

We will now show how the proof of this statement is developed, both
in the traditional procedural style of the HOL Light system,
as well in the synthesis between the procedural and declarative proof
styles that we propose in the paper.

\subsection*{The example using the procedural proof style of HOL Light}

\noindent
Traditionally, one develops the proof of a lemma in HOL Light in
an interactive session.
However, the exact commands from that session are not what is put
in the formalization file.
We now first show the session, and then the proof as it is written
in the file.

The session for this lemma consists of six commands, with after
each command output from the system:

\xmedskip
\begin{alltt}\small
# \fbox{g `!n. nsum(1..n) ({\lam}i. i) = (n*(n + 1)) DIV 2`;;\treturn\tstrut}\smallskip
val it : goalstack = 1 subgoal (1 total)

`!n. nsum (1..n) ({\lam}i. i) = (n * (n + 1)) DIV 2`

# \fbox{e INDUCT_TAC;;\treturn\tstrut}\label{tactic}\smallskip
val it : goalstack = 2 subgoals (2 total)

  0 [`nsum (1..n) ({\lam}i. i) = (n * (n + 1)) DIV 2`]

`nsum (1..SUC n) ({\lam}i. i) = (SUC n * (SUC n + 1)) DIV 2`

`nsum (1..0) ({\lam}i. i) = (0 * (0 + 1)) DIV 2`

# \fbox{e (ASM_REWRITE_TAC[NSUM_CLAUSES_NUMSEG]);;\treturn\tstrut}\smallskip
val it : goalstack = 1 subgoal (2 total)

`(if 1 = 0 then 0 else 0) = (0 * (0 + 1)) DIV 2`

# \fbox{e ARITH_TAC;;\treturn\tstrut}\smallskip
val it : goalstack = 1 subgoal (1 total)

  0 [`nsum (1..n) ({\lam}i. i) = (n * (n + 1)) DIV 2`]

`nsum (1..SUC n) ({\lam}i. i) = (SUC n * (SUC n + 1)) DIV 2`

# \fbox{e (ASM_REWRITE_TAC[NSUM_CLAUSES_NUMSEG]);;\treturn\tstrut}\smallskip
val it : goalstack = 1 subgoal (1 total)

  0 [`nsum (1..n) ({\lam}i. i) = (n * (n + 1)) DIV 2`]

`(if 1 <= SUC n then (n * (n + 1)) DIV 2 + SUC n else (n * (n + 1)) DIV 2) =\toolong
 (SUC n * (SUC n + 1)) DIV 2`

# \fbox{e ARITH_TAC;;\treturn\tstrut}\smallskip
val it : goalstack = No subgoals
\end{alltt}
\xmedskip

\noindent
The session starts with a \texttt{g} command that sets the goal to be proved,
and then executes five tactics using the \texttt{e} command.
Each time after a tactic, the system presents the subgoals that the
tactic produced, where the assumptions from which the subgoal
has to be proved are numbered (from 0), and the statement to be
proved is unnumbered.
If there are multiple subgoals produced (as is the case with
\texttt{INDUCT\char`\_TAC}), the first goal to be worked on is printed
last.

The proof as it appears in the formalization file uses
a more compact representation of these commands.
It consists of just three lines, containing the name of the
lemma, the statement, and the sequence of tactics separated
by \texttt{THEN}s:
\xmedskip
\begin{flushleft}
\fbox{\parbox{428.8pt}{%
\small
\texttt{\tstrut let ARITHMETIC\char`\_SUM = prove} \\
\texttt{\tstrut\ (`!n.\ nsum(1..n) ({\lam}i.\ i) = (n*(n + 1)) DIV 2`,} \\
\texttt{\tstrut\ \ INDUCT\char`\_TAC THEN ASM\char`\_REWRITE\char`\_TAC[NSUM\char`\_CLAUSES\char`\_NUMSEG] THEN ARITH\char`\_TAC);;}
\vspace{1pt}%
}}
\end{flushleft}
\xmedskip
\noindent
This is the customary shape of lemmas in a HOL Light formalization.

\subsection*{The example using the \texttt{miz3} proof style}

\noindent
We will now show how the same proof is developed using the
\texttt{miz3} interface.
This is a synthesis of the procedural and declarative proof
styles,
but more specifically it is a close synthesis of the HOL Light
and Mizar proof styles.
For example, like in Mizar the system will modify the file being
worked on by putting error messages inside that file.
Also, the syntax of the \texttt{miz3} proofs (explained in detail
in Section~\ref{language} below) is a direct hybrid of the syntax
of Mizar and HOL Light: the proof steps are written using
Mizar syntax, but the formulas and types in those
proof use HOL Light syntax.

There are three ways to process a \texttt{miz3} proof.
First, one can just give the proof text as a command to the
OCaml interpreter running HOL Light.
The parser has been modified to recognize the following convention:
\xmedskip
\begin{center}
\begin{tabular}{cc}
\emph{quotation style} & \emph{is parsed as} \\
\noalign{\smallskip}
\hline
\noalign{\smallskip}
\texttt{ `}\hspace{2pt}\dots\texttt{`} & a term \\
\texttt{`:}\hspace{2pt}\dots\texttt{`} & a type \\
\texttt{`;}\hspace{2pt}\dots\texttt{`} & a proof
\end{tabular}
\end{center}
\xmedskip
Second, one can put the proof (without the backquotes and semicolon)
in a file with suffix \texttt{.mz3} and check that using the
checking program \texttt{miz3}.
Third, one can use the interface from the \texttt{vi} editor.
The third style is the one we will explain in the rest of this section.
For all three interaction styles to work, a HOL Light session with
the \texttt{miz3} code loaded has to be running, as a `server'.
One runs a HOL Light session in the \texttt{miz3} source directory,
and in it executes the command:

\begingroup
\def\output{$\langle\mbox{\it many lines of output}\rangle$}
\xmedskip
\begin{alltt}\small
# \fbox{#use "miz3.ml";;\treturn\tstrut}\smallskip
\output
\end{alltt}
\xmedskip
\endgroup

\noindent
Once a HOL Light session is running with \texttt{miz3} loaded,
one can have it check \texttt{miz3} proofs from a \texttt{vi}
editor session that is running in a different window (there does not even have to be a file).
When typing two keystrokes,
\emph{control-}\texttt{C} and then \emph{return},
the part of the file where the cursor is, in between two empty lines, is checked.
We will denote these keystrokes by:
\xmedskip
\begin{center}
\fbox{\creturn\tstrut}
\end{center}
\xmedskip
\noindent
If there are errors in the checked part of the file, appropriate error messages
are inserted.
For example, an unfinished proof with errors messages
inserted might look like:
\xmedskip
\begin{alltt}\small
let ARITHMETIC_SUM = thm `;
  !n. nsum(1..n) ({\lam}i. i) = (n*(n + 1)) DIV 2
  proof
    nsum(1..0) ({\lam}i. i) = 0 [1];
::                            #2
:: 2: inference time-out
    now let n be num;
      assume nsum(1..n) ({\lam}i. i) = (n*(n + 1)) DIV 2;
      thus nsum(1..SUC n) ({\lam}i. i) = (SUC n*(SUC n + 1)) DIV 2;
::                                                           #2
    end;
  qed by INDUCT_TAC,1;
::                   #1
:: 1: inference error
`;;
\end{alltt}
\xmedskip

\noindent
The first error is caused by the lemma \texttt{NSUM\char`\_CLAUSES\char`\_NUMSEG} not being referenced,
the second error has the same reason but
is also caused by the proof automation of the system not being strong enough,
and the third error is caused by the base case being wrong:
it should not be\texttt{ }\dots\ \texttt{=} \texttt{0} but\texttt{ }\dots\ \texttt{=} \texttt{(0*(0} \texttt{+} \texttt{1))} \texttt{DIV} \texttt{2}.

If one manually writes a proof for the lemma, checking for errors
all the time and fixing them until no errors remain (this is
the style of working with the Mizar system), one ends up with a proof like:

\xmedskip
\begin{alltt}\small\label{example}
let ARITHMETIC_SUM = thm `;
  !n. nsum(1..n) ({\lam}i. i) = (n*(n + 1)) DIV 2
  proof
    nsum(1..0) ({\lam}i. i) = 0 by NSUM_CLAUSES_NUMSEG;
      .= (0*(0 + 1)) DIV 2 [1];
    now let n be num;
      assume nsum(1..n) ({\lam}i. i) = (n*(n + 1)) DIV 2 [2];
      1 <= SUC n;
      nsum(1..SUC n) ({\lam}i. i) = (n*(n + 1)) DIV 2 + SUC n
        by NSUM_CLAUSES_NUMSEG,2;
      thus .= ((SUC n)*(SUC n + 1)) DIV 2;
    end;
  qed by INDUCT_TAC,1`;;
\end{alltt}
\xmedskip

\noindent
This proof is thirteen lines instead of the three that we got
with the traditional proof style.
I.e., this proof is quite a bit longer,
but not unreasonably so.
We experimented quite a bit by comparing procedural proofs to
their declarative counterparts, and our impression is that
declarative proofs are generally about twice as long as corresponding procedural
ones.
See in this respect also the statistics in Section~\ref{proofcounts} on page~\pageref{proofcounts} below.

This proof was written using \texttt{miz3} in a purely declarative style.
We now show
how one can use \texttt{miz3} in a purely procedural style,
exactly mimicking the traditional HOL Light session.
(We have the problem of how to present an interactive editing
session on paper.
We will do this by presenting various stages of the edit buffer, interspersed with
comments.
This will mean a lot of duplicated text, but hopefully it will make the
process clear.)

The standard starting point for a `procedural' \texttt{miz3} session is:

\xmedskip
\begin{alltt}\small
let  = thm `;
  
  proof
  qed by #;
`;;
\end{alltt}
\xmedskip

\noindent
In the place of the two empty spaces one puts the name and statement
of the lemma to be proved:

\label{luxury}
\xmedskip
\begin{alltt}\small
let \fbox{ARITHMETIC_SUM\tstrut} = thm `;\vspace{1pt}
  \fbox{!n. nsum(1..n) ({\lam}i. i) = (n*(n + 1)) DIV 2\tstrut}\smallskip
  proof
  qed by #;
`;;
\end{alltt}
\xmedskip

\noindent
When checking this, there will be \emph{no} error messages added, as the
\texttt{\char`\#} mark means that this line is a subgoal to be proved.
The \texttt{qed} step means that the statement
from the lemma has been proved, and therefore
the subgoal in this case consists of exactly that statement.

Next, one types a tactic after the \texttt{\char`\#} sign
and has the system process the file:

\xmedskip
\begin{alltt}\small
let ARITHMETIC_SUM = thm `;
  !n. nsum(1..n) ({\lam}i. i) = (n*(n + 1)) DIV 2
  proof
  qed by #\tskip\fbox{INDUCT_TAC\,\creturn\tstrut};\toolong\smallskip
`;;
\end{alltt}
\xmedskip

\noindent
The tactic will be executed, and the system will `merge' the two subgoals that are generated
(see the traditional session on page~\pageref{tactic})
into the proof, using the method from \cite{kal:wie:09}.
Also, the insertion point of the editor will be put directly after the
first \texttt{\char`\#} as indicated by the vertical bar, i.e., the editor will `jump' to the first subgoal that
is now left:

\xmedskip
\begin{alltt}\small
let ARITHMETIC_SUM = thm `;
  !n. nsum(1..n) ({\lam}i. i) = (n*(n + 1)) DIV 2
  proof
    nsum (1..0) ({\lam}i. i) = (0 * (0 + 1)) DIV 2 [1] by #\tbar;
    !n. nsum (1..n) ({\lam}i. i) = (n * (n + 1)) DIV 2
        ==> nsum (1..SUC n) ({\lam}i. i) = (SUC n * (SUC n + 1)) DIV 2 [2]
    proof
      let n be num;
      assume nsum (1..n) ({\lam}i. i) = (n * (n + 1)) DIV 2;
    qed by #;
  qed by INDUCT_TAC from 1,2;
`;;
\end{alltt}
\xmedskip

\noindent
There is quite some text already, but most of it is generated
and not typed by the user.
Next one enters the second tactic, and has the system process it:

\xmedskip
\begingroup
\makeatletter
\def\flbox#1{\leavevmode\setbox\@tempboxa\hbox{#1}\@tempdima\fboxrule
    \advance\@tempdima \fboxsep \advance\@tempdima \dp\@tempboxa
   \hbox{\lower \@tempdima\hbox
  {\vbox{\hrule \@height \fboxrule
          \hbox{\vrule \@width \fboxrule \hskip\fboxsep
          \vbox{\vskip\fboxsep \box\@tempboxa\vskip\fboxsep}%
                 }%
                 \hrule \@height \fboxrule}}}}
\def\frbox#1{\leavevmode\setbox\@tempboxa\hbox{#1}\@tempdima\fboxrule
    \advance\@tempdima \fboxsep \advance\@tempdima \dp\@tempboxa
   \hbox{\lower \@tempdima\hbox
  {\vbox{\hrule \@height \fboxrule
          \hbox{%
          \vbox{\vskip\fboxsep \box\@tempboxa\vskip\fboxsep}\hskip
                 \fboxsep\vrule \@width \fboxrule}%
                 \hrule \@height \fboxrule}}}}
\makeatother
\begin{alltt}\small
let ARITHMETIC_SUM = thm `;
  !n. nsum(1..n) ({\lam}i. i) = (n*(n + 1)) DIV 2
  proof
    nsum (1..0) ({\lam}i. i) = (0 * (0 + 1)) DIV 2 [1] by #\tskip\flbox{REWRITE_TAC,NSUM_CLAUSES_NU\tstrut}
\frbox{MSEG\,\creturn\tstrut};\toolong\smallskip
    !n. nsum (1..n) ({\lam}i. i) = (n * (n + 1)) DIV 2
        ==> nsum (1..SUC n) ({\lam}i. i) = (SUC n * (SUC n + 1)) DIV 2 [2]
    proof
      let n be num;
      assume nsum (1..n) ({\lam}i. i) = (n * (n + 1)) DIV 2;
    qed by #;
  qed by INDUCT_TAC from 1,2;
`;;
\end{alltt}
\endgroup
\xmedskip

\noindent
This leads to a new subgoal, again exactly matching the one from
the traditional HOL Light session:

\xmedskip
\begin{alltt}\small
let ARITHMETIC_SUM = thm `;
  !n. nsum(1..n) ({\lam}i. i) = (n*(n + 1)) DIV 2
  proof
    (if 1 = 0 then 0 else 0) = (0 * (0 + 1)) DIV 2 [1] by #\tbar;
    nsum (1..0) ({\lam}i. i) = (0 * (0 + 1)) DIV 2 [2]
      by REWRITE_TAC,NSUM_CLAUSES_NUMSEG from 1;
    !n. nsum (1..n) ({\lam}i. i) = (n * (n + 1)) DIV 2
        ==> nsum (1..SUC n) ({\lam}i. i) = (SUC n * (SUC n + 1)) DIV 2 [3]
    proof
      let n be num;
      assume nsum (1..n) ({\lam}i. i) = (n * (n + 1)) DIV 2;
    qed by #;
  qed by INDUCT_TAC from 2,3;
`;;
\end{alltt}
\xmedskip

\noindent
Note that the system automatically wrapped the line with the long tactic.

If one continues like this by putting in three more tactics,
one gets the declarative counterpart of the procedural proof:

\xmedskip
\begin{alltt}\small
let ARITHMETIC_SUM = thm `;
  !n. nsum(1..n) ({\lam}i. i) = (n*(n + 1)) DIV 2
  proof
    (if 1 = 0 then 0 else 0) = (0 * (0 + 1)) DIV 2 [1] by ARITH_TAC;
    nsum (1..0) ({\lam}i. i) = (0 * (0 + 1)) DIV 2 [2]
      by REWRITE_TAC,NSUM_CLAUSES_NUMSEG from 1;
    !n. nsum (1..n) ({\lam}i. i) = (n * (n + 1)) DIV 2
        ==> nsum (1..SUC n) ({\lam}i. i) = (SUC n * (SUC n + 1)) DIV 2 [3]
    proof
      let n be num;
      assume nsum (1..n) ({\lam}i. i) = (n * (n + 1)) DIV 2;
      (if 1 <= SUC n
       then nsum (1..n) ({\lam}i. i) + SUC n
       else nsum (1..n) ({\lam}i. i)) =
      (SUC n * (SUC n + 1)) DIV 2 [4] by ARITH_TAC;
    qed by REWRITE_TAC,NSUM_CLAUSES_NUMSEG from 4;
  qed by INDUCT_TAC from 2,3;
`;;
\end{alltt}
\xmedskip

\noindent
This is
not a proof a human would have written, but it \emph{exactly}
matches the traditional HOL Light session.
Furthermore, the characters that one needs to type to get it
are almost exactly the same as in that session.

Now the point of this paper is that one does \emph{not} need
to choose between these two ways of working.
One can work on a proof in a combination, both freely editing
it and rechecking things while going, but also by executing
tactics at various places and using the new proof steps that these tactics
produce.

Note that \texttt{miz3} just replaces the \emph{proof} part of
the HOL Light mathematical language.
The rest of HOL Light's features like definitions and
the implementation of proof automation are unchanged.
Therefore, for convenience the \fbox{\creturn\tstrut} command will interpret
text being processed as straight OCaml code if it
does not have the shape
\begingroup
\def\name{$\langle\mbox{ident}\rangle$}
\def\rest{$\dots$}
\xmedskip
\begin{alltt}
  \dots thm `;
    \dots
  `;;
\end{alltt}
\xmedskip
\endgroup
\noindent
where the \texttt{`;} has to be at the end of the first line
of the block.
This means that there is no need to enter text directly in the HOL Light session,
one can fully work from the \texttt{vi} interface.
However, working in the HOL Light session itself is also possible.

\section{The \texttt{miz3} proof language}\label{language}

\noindent
The three most common proof systems for first order predicate
logic are natural deduction, sequent calculus and Hilbert-style logic.
Of these three, natural deduction corresponds most closely with
everyday mathematical reasoning.
The two main proof systems for natural deduction are
Ja\'skowski/Fitch- and Gentzen-style deduction \cite{pel:00}.
Of these, the first is the easiest to use for actual proofs.
An example of a proof in a Ja\'skowski/Fitch-style natural deduction
proof system (where we use the syntactic conventions of \cite{hut:rya:04}) is:

\begingroup
\setlength{\tabcolsep}{.5em}
\def\lstrut{{\normalsize\strut}}
\def\Lstrut{{\large\strut}}
\def\skipforward{\noalign{\smallskip}}
\def\mbar{\multicolumn{1}{|l}{}}
\xmedskip
\begin{center}
\small
\begin{tabular}{rlllrllllll}
\lstrut 1 &&&&& $(\exists x\,\neg P(x)) \lor \neg(\exists x\,\neg P(x))$ & {\scriptsize LEM} \\
\cline{2-10}
\lstrut 2 & \mbar &&&& $\exists x\,\neg P(x)$ & assumption &&&& \mbar \\
\cline{4-8}
\lstrut 3 & \mbar && \mbar & $x$ & $\neg P(x)$ & assumption && \mbar && \mbar \\
\cline{5-7}
\lstrut 4 & \mbar && \mbar & \mbar & $P(x)$ & assumption & \mbar & \mbar && \mbar \\
\lstrut 5 & \mbar && \mbar & \mbar & $\bot$ & $\neg E$ 3,4 & \mbar & \mbar && \mbar \\
\lstrut 6 & \mbar && \mbar & \mbar & $\forall y\, P(y)$ & $\bot E$ 5 & \mbar & \mbar && \mbar \\
\cline{5-7}
\lstrut 7 & \mbar && \mbar && $P(x) \To \forall y\, P(y)$ & ${\To}I$ 4--6 && \mbar && \mbar \\
\lstrut 8 & \mbar && \mbar && $\exists x (P(x) \To \forall y\, P(y))$ & $\exists I$ 7 && \mbar && \mbar \\
\cline{4-8}
\lstrut 9 & \mbar &&&& $\exists x (P(x) \To \forall y\, P(y))$ & $\exists E$ 2,3--8 &&&& \mbar \\
\cline{2-10}
&&&&&& \\
\noalign{\vspace{-\medskipamount}}
\cline{2-10}
\lstrut 10 & \mbar &&&& $\neg(\exists x\,\neg P(x))$ & assumption &&&& \mbar \\
\cline{3-9}
\lstrut 11 & \mbar & \mbar &&& $P(a)$ & assumption &&& \mbar & \mbar \\
\cline{4-8}
\lstrut 12 & \mbar & \mbar & \mbar & $y$ & $P(y) \lor \neg P(y)$ & {\scriptsize LEM} && \mbar & \mbar & \mbar \\
\cline{5-7}
\lstrut 13 & \mbar & \mbar & \mbar & \mbar & $P(y)$ & assumption & \mbar & \mbar & \mbar & \mbar \\
\cline{5-7}
& \mbar & \mbar & \mbar &&&&& \mbar & \mbar & \mbar \\
\noalign{\vspace{-\medskipamount}}
\cline{5-7}
\lstrut 14 & \mbar & \mbar & \mbar & \mbar & $\neg P(y)$ & assumption & \mbar & \mbar & \mbar & \mbar \\
\lstrut 15 & \mbar & \mbar & \mbar & \mbar & $\exists x\, \neg P(x)$ & $\exists I$ 14 & \mbar & \mbar & \mbar & \mbar \\
\lstrut 16 & \mbar & \mbar & \mbar & \mbar & $\bot$ & $\neg E$ 10,15 & \mbar & \mbar & \mbar & \mbar \\
\lstrut 17 & \mbar & \mbar & \mbar & \mbar & $P(y)$ & $\bot E$ 16 & \mbar & \mbar & \mbar & \mbar \\
\cline{5-7}
\lstrut 18 & \mbar & \mbar & \mbar && $P(y)$ & $\lor E$ 12,13,14--17 && \mbar & \mbar & \mbar \\
\cline{4-8}
\lstrut 19 & \mbar & \mbar &&& $\forall y\, P(y)$ & $\forall I$ 12--18 &&& \mbar & \mbar \\
\cline{3-9}
\lstrut 20 & \mbar &&&& $P(a) \To \forall y\, P(y)$ & ${\To}I$ 11--19 &&&& \mbar \\
\lstrut 21 & \mbar &&&& $\exists x (P(x) \To \forall y\, P(y))$ & $\exists I$ 20 &&&& \mbar \\
\cline{2-10}
\lstrut 22 &&&&& $\exists x (P(x) \To \forall y\, P(y))$ & $\lor E$ 1,2--9,10--21
\end{tabular}
\end{center}
\xmedskip
\xmedskip
\endgroup

\noindent
This is a proof of Smullyan's `{drinker's principle}' \cite{smu:90}.

Most declarative systems have essentially the same proof language
(although the \emph{explanation} of the
{meaning} of the elements of such a language can differ significantly
between systems \cite{wen:wie:02}.)
In other words, declarative interactive theorem provers have proof
languages that are much more similar to each other than the languages of procedural systems
are.
It turns out that the languages of declarative systems are all close to Ja\'skowski/Fitch-style
natural deduction.

The \texttt{miz3} proof language is in particular almost identical to the language of
the Mizar system \cite{gra:kor:nau:10}.
The natural deduction example that we just gave becomes in the
syntax of the \texttt{miz3} language:

\begingroup
\def\\{\char`\\}
\xmedskip
\begin{alltt}\small
  let DRINKER = thm `;
    let P be A->bool;
    thus ?x. P x ==> !y. P y [22]
    proof
      (?x. ~P x) \\/ ~(?x. ~P x) [1];
      cases by 1;
      suppose ?x. ~P x [2];
        consider x such that ~P x [3] by 2;
        take x;
        assume P x [4];
        F [5] by 3,4;
        thus !y. P y [6] by 5;
      end;
\end{alltt}
\begin{alltt}\small
      suppose ~(?x. ~P x) [10];
        consider a being A such that T;
        take a;
        assume P a [11];
        let y be A;
        P y \\/ ~P y [12];
        cases by 12;
        suppose P y [13];
          thus P y by 13;
        end;
        suppose ~P y [14];
          ?x. ~P x [15] by 14;
          F [16] by 10,15;
          thus P y [17] by 16;
        end;
      end;
    end`;;
\end{alltt}
\xmedskip
\xmedskip
\endgroup

\noindent
(The extra
`\texttt{consider} \texttt{a} \texttt{being} \texttt{A} \texttt{such} \texttt{that} \texttt{T;}' step
is needed because in \texttt{miz3}
variables have to be introduced before they can be used.
The Ja\'skowski/Fitch-style natural deduction proof can
use the free variable $a$
without introducing it,
but in \texttt{miz3} this is not allowed.
The step should be read as `consider an object \texttt{a} of type \texttt{A} such that true holds'.
It is accepted by the system because in HOL all types are non-empty.
This non-emptiness is indeed essential for
the drinker's principle to be provable.)

This example shows the correspondence between \texttt{miz3} proof steps
and natural deduction rules:

\xmedskip
\begin{center}
\begin{tabular}{cc}
\emph{deduction rule} & \emph{proof step} \\
\noalign{\smallskip}
\hline
\noalign{\smallskip}
assumption & \texttt{thus} \\
\noalign{\medskip}
${\land}I$ & \texttt{thus} \\
${\To}I$ & \texttt{assume} \\
${\neg}I$ & \texttt{assume} \\
${\forall}I$ & \texttt{let} \\
${\exists}I$ & \texttt{take} \\
\noalign{\medskip}
${\lor}E$ & \texttt{cases}/\texttt{suppose} \\
${\exists}E$ & \texttt{consider}
\end{tabular}
\end{center}
\xmedskip
\xmedskip
\xmedskip

\noindent
The other natural deduction rules, like ${\land}E$, ${\Rightarrow}E$, ${\neg}E$, ${\forall}E$ and ${\lor}I$, are all subsumed by the
inference checker `\texttt{by}'.
Actually, the ${\land}I$ and ${\exists}I$ rules are subsumed by \texttt{by} as well.
The \texttt{thus} and \texttt{take} steps are more of a backward step than a direct
counterpart to the
${\land}I$ and ${\exists}I$ rules.

We now present the meaning of the \texttt{miz3} language.
It is almost exactly the same as that
of the Mizar language \cite{gra:kor:nau:10}.
At every step in a proof there is a designated statement called the \texttt{thesis}.
This is the statement that is being proved (the \emph{goal} of a procedural prover).
Subproofs have their own local \texttt{thesis}.
Now steps can add extra variables and statements to the proof context,
but can also change the \texttt{thesis}.
If that happens the step is called
a \emph{skeleton} step.
Here is a table that summarizes the main \texttt{miz3} proof steps:

\xmedskip
\begin{center}
$\hspace{-5em}$\begin{tabular}{lcccccc}
\emph{proof step} & \emph{statement} & \emph{new proof} & \emph{referable} & \emph{old} & \emph{new} & \emph{skeleton} \\
 & \emph{to be justified} & \emph{variables} & \emph{statement} & \texttt{thesis} & \texttt{thesis} & \emph{step?} \\
\noalign{\smallskip}
\hline
\noalign{\medskip}
$\phi$\texttt{ by }\dots\texttt{;} & $\phi$ & & $\phi$ & $\psi$ & $\psi$ & $-$ \\
\noalign{\medskip}
\texttt{let }$x_1$ \dots\ $x_n${\tt\ be }$\tau$\texttt{;}$\hspace{-0em}$ & & $x_1$ \dots\ $x_n$ & & $\forall x_1 \dots\forall x_n\,\psi$ & $\psi$ & $+$ \\
\noalign{\medskip}
\texttt{assume }$\phi$\texttt{;} & & & $\phi$ & $\phi \Rightarrow \psi$ & $\psi$ & $+$ \\
\noalign{\medskip}
\texttt{assume }$\phi$\texttt{;} & & & $\phi$ & $\neg\phi$ & $\bot$ & $+$ \\
\noalign{\medskip}
\texttt{assume }$\neg\phi$\texttt{;} & & & $\neg\phi$ & $\phi$ & $\bot$ & $+$ \\
\noalign{\medskip}
\texttt{thus }$\phi$\texttt{ by }\dots\texttt{;} & $\phi$ & & $\phi$  & $\phi \land \psi$ & $\psi$ & $+$ \\
\noalign{\medskip}
\texttt{thus }$\phi$\texttt{ by }\dots\texttt{;} & $\phi$ & & $\phi$ & $\phi$ & $\top$ & $+$ \\
\noalign{\medskip}
\texttt{qed by }\dots\texttt{;} & $\phi$ & & & $\phi$ & & $+$ \\
\noalign{\medskip}
\texttt{take }$t$\texttt{;} & & & & $\exists x\,\psi$ & $\psi[x:=t]$ & $+$ \\
\noalign{\medskip}
\texttt{.= }$t_{i + 1}$\texttt{ by }\dots\texttt{;} & $t_i = t_{i + 1}$ & & $t_1 = t_{i + 1}$ & $\psi$ & $\psi$ & $-$ \\
\noalign{\medskip}
\texttt{set }$x$\texttt{ = }$t$\texttt{;} & & $x$ & $x = t$ & $\psi$ & $\psi$ & $-$ \\
\noalign{\smallskip}
\hline
\noalign{\smallskip}
\texttt{consider }$x_1$ \dots\ $x_n\hspace{-0em}$ \\
$\hspace{1em}$ \texttt{such that} \\
$\hspace{1em}$ $\phi$\texttt{ by }\dots\texttt{;} & $\exists x_1\dots\exists x_n\, \phi$ & $x_1$ \dots\ $x_n$ & $\phi$ & $\psi$ & $\psi$ & $-$ \\
\noalign{\smallskip}
\hline
\noalign{\smallskip}
\texttt{cases by }\dots\texttt{;} & $\phi_1 \lor \dots \lor \phi_n$ & & & $\psi$ \\
\dots \\
\texttt{suppose }$\phi_i$\texttt{;} & & & $\phi_i$ & & $\psi$ & $+$ \\
$\hspace{1em}$ \dots \\
\texttt{end;} \\
\dots \\
\noalign{\smallskip}
\hline
\end{tabular}$\hspace{-5em}$
\end{center}
\xmedskip
\xmedskip
\noindent
This then is the basic \texttt{miz3} proof language.
The \texttt{by} justifications should contain sufficient references for
the automation to prove the statement in the second column.
The third and fourth columns give the variables and statements that
are added to the proof context.
The fifth and sixth columns give the \texttt{thesis} before and
after the step, and the seventh column
indicates whether the step is a skeleton step or not.

In the case of a \texttt{cases}, every \texttt{suppose} branch
inherits the \texttt{thesis} that held at the \texttt{cases}, as indicated in the table.
However, as one cannot remove a \texttt{suppose} without destroying
the skeleton of the proof, it still counts as a skeleton step.

\begin{figure}
\vspace{2em}
\begin{eqnarray*}
\xtoolong\notion{lemma} &::=& \overbrace{\big(\key{let}\terminal{ident}\literal{=}\big)\!\strut^?\,\key{thm}\literal{`;}
\overbrace{\terminal{formula}\notion{proof}\literal{;}^?}^{\mbox{\small \texttt{miz3\vrule height 8.5pt depth 2pt width 0pt}}}
\literal{`}\literal{;;}}^{\mbox{\small HOL Light}}\xtoolong \\
\noalign{\medskip}
\notion{proof} &::=& \notion{by refs} \\
&\alter&
\key{proof}\notion{proof step}\!^*\,(\key{end}\,|\,\key{qed}\notion{by refs}) \\
\noalign{\medskip}
\xtoolong\notion{proof step} &::=& \terminal{formula}\notion{labels}\notion{proof}\literal{;} \\
&\alter& \key{let}\terminal{ident}\!^+\,\key{be}\terminal{type}\literal{;} \\
&\alter& \key{assume}\terminal{formula}\notion{labels}\,\literal{;} \\
&\alter& \key{thus}\terminal{formula}\notion{labels}\notion{proof}\literal{;} \\
&\alter& \key{take}\terminal{term}\literal{;} \\
&\alter& \key{consider}\terminal{ident}\!^+\,(\key{being}\notion{type})^?\,\key{such}\key{that} \\
&& \quad \terminal{formula}\notion{labels}
\notion{proof}\literal{;}\xtoolong \\
&\alter& \key{cases}\notion{by refs}\literal{;} \\
&& \big(\key{suppose}\terminal{formula}\notion{labels}\,\literal{;} \\
&& \quad \notion{proof step}\!^*\, (\key{end}\,|\,\key{qed}\notion{by refs})\,\literal{;}\big)^* \\
&\alter& \key{now}\notion{labels}\notion{proof step}\!^+\,\key{end}\literal{;} \\
&\alter& \literal{.=}\terminal{term}\notion{labels}\notion{proof}\literal{;} \\
&\alter& \key{set}\terminal{ident}\literal{=}\terminal{term}\notion{labels}\,\literal{;} \\
&\alter& \key{exec}\terminal{tactic}\literal{;} \\
\noalign{\medskip}
\notion{labels} &::=& \big(\literal{[}\terminal{ident}\literal{]}\big)^* \\
\noalign{\medskip}
\xtoolong\notion{by refs} &::=& \big(\key{by}\notion{ref}(\literal{,}\notion{ref})^*\big)^?\,\big(\key{from}\notion{ref}(\literal{,}\notion{ref})^*\big)^? \\
\noalign{\medskip}
\notion{ref} &::=& \literal{-} \;|\; \terminal{ident}\;|\; \underbrace{\strut\terminal{{thm}}\;|\; \terminal{{tactic}}\;|\; \terminal{{conv}}}_{\mbox{\small\strut OCaml}} \\
\end{eqnarray*}
\caption{The full \texttt{miz3} grammar}
\label{grammar}
\vspace{2em}
\end{figure}

The \texttt{.=} construction is used for \emph{iterated equalities}.
Not indicated in the table (for lack of space) is that the step directly before the
\texttt{.=}\,
should prove a statement of the shape $t_1 = t_i$.
This then determines the terms $t_1$ and $t_i$ in the table.
Using \texttt{.=} reasoning, chained equalities of the shape
$t_1 = t_2 = t_3 = \dots = t_n$
can be coded as
\xmedskip
\begin{flushleft}\tt
\ \ $t_1$ = $t_2$ by {\rm\dots}; \\
\ \ \phantom{$t_1$} \llap{\tt .}= $t_3$ by {\rm\dots}; \\
\ \ \phantom{$t_1$} \llap{\rlap{\rm\dots}\phantom{\tt .}} \\
\ \ \phantom{$t_1$} \llap{\tt .}= $t_n$ by {\rm\dots};
\end{flushleft}
\xmedskip
\noindent
\looseness=-1
The \texttt{now} syntax can be used when the statement that is
being proved can be inferred from the skeleton steps
in the proof (which is generally the case).
In that situation
\xmedskip
\begingroup
\def\form{$\phi$}
\def\xdots{\dots}
\begin{alltt}
  \form proof \xdots end;
\end{alltt}
\xmedskip
\noindent
can be abbreviated as
\xmedskip
\begin{alltt}
  now \xdots end;
\end{alltt}
\endgroup
\xmedskip
\noindent
Finally the \texttt{-} label refers to the statement that was introduced most recently to the proof context,
and the \texttt{exec} step can be used to transform the \texttt{thesis} using an arbitrary tactic in the style of \cite{wie:01}.

The \texttt{miz3} language differs in some respects from the real Mizar language:
\begin{iteMize}{$\bullet$}
\item
The statements, terms and types use HOL Light syntax instead of Mizar 
syntax.
In particular the rich type system of Mizar \cite{wie:07:2} is not available.

\item
The labels are behind the statements in brackets, instead of
in front with a colon.

\item
There is no \texttt{then}.
The label \texttt{-} refers to the previous statement.
Also, the last `\texttt{horizon}' statements are visible without
reference, where \texttt{horizon} is a variable of the
\texttt{miz3} server that is usually set to \texttt{1}.
If one sets \texttt{horizon} \texttt{:=} \texttt{-1;;} all statements local
to the proof that are in scope are used in inference checking
without any references.
(This makes the proofs look nice and checking very slow.)

\item
Some keywords are slightly different:
see Figure~\ref{grammar} for the exact grammar.
This grammar can be parsed with \texttt{yacc} without
any shift/reduce conflicts.
In that sense it is more straightforward than the grammar used by the real Mizar \cite{cai:gow:04}.
The most notable changes are that
there is a keyword \texttt{qed} as an abbreviation of `\texttt{thus thesis; end}',
and that in iterated equalities {each} step has
a terminating semicolon instead of just the last one.

\item
The \texttt{from} keyword has a different meaning than in Mizar
(see the explanation below of how tactics in justifications function.)

\end{iteMize}

\noindent
In a justification of a proof step an arbitrary HOL Light tactic (either of
type \texttt{tactic} or of type \texttt{thm list} \texttt{->} \texttt{tactic}) can be given.
If this tactic is absent, the default tactic \texttt{default\char`\_{\penalty 100}prover}, which is another variable of the \texttt{miz3} server, is used.
This first tries \texttt{HOL\char`\_BY},
an equivalent of the Mizar automated theorem prover
by John Harrison, and if that fails runs some decision procedures, as well as the
HOL Light first order automated theorem prover \texttt{MESON} \cite{har:96:1}.

If a tactic is explicitly present in a step,
a goalstate is created with the statement that needs to be justified as the goal, and the statements referred to in the \texttt{by}
part of the justification as the assumptions.
In this goalstate the tactic is executed with the \texttt{by} part
statements as arguments.
Finally it is checked whether the subgoals obtained that way
are all present in the union of the \texttt{by} and \texttt{from}
parts of the justification.
A tactic to connect everything in the style
of \cite{wie:01} is executed, which will fail if this is not the case.

\section{The implementation of the \texttt{miz3} proof interface}\label{implementation}

\noindent
We now present a low-level description of the implementation of the
\texttt{miz3} interface.
This might seem to be about irrelevant details, but if one leaves out
enhancements like the caches and time-outs, the system becomes unusable
for serious work.
Readers only interested in an abstract description of the
\texttt{miz3} proof style, or in using the system instead of understanding
how it is organized on the inside, can safely skip this section.

The \texttt{miz3} interface is both a prototype (in the sense that
it has no real users yet) as well as a quite
usable system (in the sense that it is a quite usable proof language for
the quite usable HOL Light system).
The \texttt{miz3} interface is rather light weight, although
in its current implementation
it needs the infrastructure of a Unix system.

The full source of the \texttt{miz3} system consists of five files:
\xmedskip
\begin{center}
\begin{tabular}{lrll}
\texttt{miz3.ml} & 1,903 lines && the OCaml program to be loaded in HOL Light \\
\texttt{miz3\char`\_of\char`\_hol.ml} & 237 lines && (see Section~\ref{automatic} below) \\
\texttt{bin/miz3} & 28 lines && the command line batch checker \\
\texttt{bin/miz3f} & 40 lines && a filter version of the command line checker \\
\texttt{exrc} & 9 lines && the `\texttt{vi} mode' for \texttt{miz3}
\end{tabular}
\end{center}
\xmedskip

\noindent
The communication between the \texttt{vi} session and HOL Light
is initiated by \texttt{vi} sending a Unix signal to HOL Light,
after which the signal handler of the HOL Light session does all the work.
This means that the HOL Light session when not doing a \texttt{miz3}
check is {not} actively `waiting' for commands, and that if it is
running some other code, then that code will be interrupted to do the \texttt{miz3}
check.
The communication between \texttt{vi} and HOL Light is through
files, of which some have hard coded special names in order for both
sides to be able to find them.

Specifically, when checking a proof from the \texttt{vi} interface by
typing
\xmedskip
\begin{center}
\fbox{\creturn\tstrut}
\end{center}
\xmedskip
the following steps happen:
\begin{iteMize}{(1)}
\item
The relevant part of the buffer is selected using the \texttt{vi}
commands \texttt{\char`\{} and \texttt{\char`\}}.

\item
This part of the buffer is filtered through the perl script \texttt{miz3f}.

\item
This script writes this to a temporary file in \texttt{/tmp},
and runs the shell script \texttt{miz3} on that file.

\item
This then writes the name of its input file in the file with the special name \texttt{/tmp/{\penalty 100}miz3\char`\_{\penalty 100}filename},
looks up the process id of the HOL Light session running the \texttt{miz3}
server in \texttt{/tmp/miz3\char`\_pid}, and sends that process the \texttt{USR2}
signal.

\item
The signal handler of the HOL Light session now finds the file it needs
to check, and parses it into a data structure of OCaml type
\texttt{step} \texttt{list}.
In this data structure the full input is stored in small pieces.

\item
Next, the server calls the function \texttt{check\char`\_proof} on this,
which returns another \texttt{step} \texttt{list}, this time with errors
marked, and possibly grown by running tactics after the \texttt{\char`\#}
symbol.

\item
This result now is printed back into the file in \texttt{/tmp},
after which the filter puts it back into the edit buffer.

\end{iteMize}

\noindent
In other words the \emph{full} proof is processed \emph{every} time a check is done.
To make this acceptably fast, there are two caches.
The first cache holds inferences that have already been checked,
to prevent the checker from having to run all tactics every time.
The second cache holds the OCaml objects associated with
the elements in the \texttt{by} justifications.
These are calculated using the OCaml functions
\texttt{Toploop.{\penalty 100}parse\char`\_toplevel\char`\_phrase} and
\texttt{Toploop.execute\char`\_phrase} (which together are the
OCaml equivalent of Lisp's \texttt{eval} function).

We do not want to restrict users to `special' tactics that
are designed to always terminate in a reasonable time.
This means that while working on a proof sometimes tactics
will `hang'.
Therefore, while developing a proof, after a specified time tactics will be killed
using the \texttt{Unix.alarm} function.
This time is given by the variable \texttt{timeout}.
Of course, this makes it dependent on the specific computer used whether a proof will be accepted 
or flagged with a time-out error.
(However, a similar thing holds for any interactive theorem prover, because the memory might run out
during a check as well.)
If one wants to check a finished proof on a slow system without having
to worry about time-outs for tactics, one can set \texttt{timeout} \texttt{:=} \texttt{-1;;}
to disable time-outs.

The system will never pretty-print user input by itself,
even though it parses and reprints proofs all the time.
Even white space and comments will stay exactly the way they are.
However, if the proof is modified by `growing' it through
execution of tactics,
the system tries hard to indent and line wrap the new parts nicely.
There are a dozen parameters to direct this process, which
are listed at the start of the source file \texttt{miz3.ml}.

Although the interface currently only runs in \texttt{vi}, we took great care
to not have it be dependent on \texttt{vi} specific features.
In fact, the whole `\texttt{miz3} mode' for \texttt{vi} essentially consists of
the single line
\xmedskip
\begingroup
\def\{{\char`\{}
\def\}{\char`\}}
\begin{alltt}
:map ^C<CR> \{/.^M!\}miz3f^M/#^Ml
\end{alltt}
\endgroup
\xmedskip
\noindent
in the \texttt{.exrc} file (for \texttt{vim} users the \texttt{.vimrc} file),
which is the file that configures mappings between \texttt{vi} keystrokes and commands.
The simplicity of this interface means that porting it to other editors will be trivial.

\section{Automatically converting procedural proofs to declarative proofs}\label{automatic}

\noindent
Traditional HOL Light proofs can be mimicked in \texttt{miz3} by using the system
in the procedural style
as shown in Section~\ref{session}.
We wrote a small program that automates this
process.
Using this, \emph{any} proof from HOL Light library
can be fully automatically converted to the \texttt{miz3} language.

For example, consider the following HOL Light lemma, that
says that subtraction on natural and real numbers correspond to each other:

\xmedskip
\begin{alltt}\small
let REAL_OF_NUM_SUB = prove
 (`!m n. m <= n ==> (&n - &m = &(n - m))`,
  REPEAT GEN_TAC THEN REWRITE_TAC[LE_EXISTS] THEN
  STRIP_TAC THEN ASM_REWRITE_TAC[ADD_SUB2] THEN
  REWRITE_TAC[GSYM REAL_OF_NUM_ADD] THEN
  ONCE_REWRITE_TAC[REAL_ADD_SYM] THEN
  REWRITE_TAC[real_sub; GSYM REAL_ADD_ASSOC] THEN
  MESON_TAC[REAL_ADD_LINV; REAL_ADD_SYM; REAL_ADD_LID]);;
\end{alltt}
\xmedskip

\noindent
To be able to use the converter, it is necessary to load it:

\begingroup
\def\output{$\langle\mbox{\it many lines of output}\rangle$}
\xmedskip
\begin{alltt}\small
# \fbox{#use "miz3_of_hol.ml";;\treturn\tstrut}\smallskip
\output
\end{alltt}
\xmedskip
\endgroup
\noindent
Then one converts the proof as follows:

\xmedskip
\begin{alltt}\small
# \fbox{miz3_of_hol "/opt/src/hol_light/real.ml" "REAL_OF_NUM_SUB";;\treturn\tstrut}\smallskip
0..0..1..2..7..14..37..72..174..325..solved at 392
val it : step =
  !m n. m <= n ==> &n - &m = &(n - m) [1]
  proof
    !n m. m <= n ==> &n - &m = &(n - m) [2]
    proof
      let n m be num;
      !d. n = m + d ==> &n - &m = &(n - m) [3]
      proof
        let d be num;
        assume n = m + d [4];
        &d + &m + -- &m = &d [5]
          by MESON_TAC[REAL_ADD_LINV; REAL_ADD_SYM; REAL_ADD_LID],4;
        (&d + &m) - &m = &d [6]
          by REWRITE_TAC[real_sub; GSYM REAL_ADD_ASSOC],4 from 5;
        (&m + &d) - &m = &d [7] by ONCE_REWRITE_TAC[REAL_ADD_SYM],4 from 6;\toolong
        &(m + d) - &m = &d [8]
          by REWRITE_TAC[GSYM REAL_OF_NUM_ADD],4 from 7;
      qed by ASM_REWRITE_TAC[ADD_SUB2],4 from 8;
      (?d. n = m + d) ==> &n - &m = &(n - m) [9] by STRIP_TAC from 3;
    qed by REWRITE_TAC[LE_EXISTS] from 9;
  qed by REPEAT GEN_TAC;
\end{alltt}
\xmedskip

\noindent
The possibility to convert proofs like this does not depend on HOL Light specifics.
Any procedural system that has goals and tactics can have a \texttt{miz3} layer
on top of it and proofs can then be converted to that in exactly the same way.
We believe that this gives an approach to `integrate' mathematical libraries
between systems:

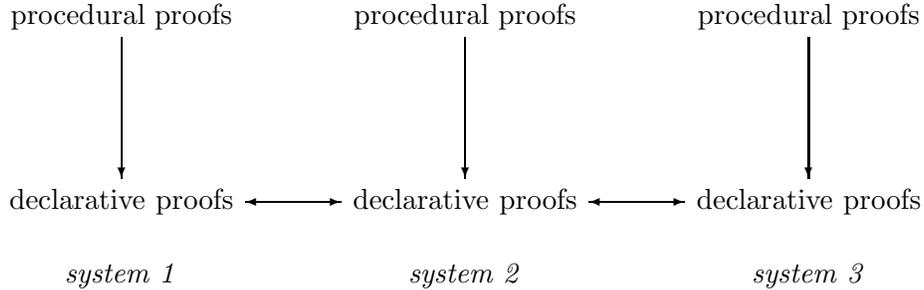
\xmedskip
\begin{center}
\begin{picture}(340,120)(0,22)
\put(40,32){\makebox(0,0){\strut \emph{system 1}}}
\put(170,32){\makebox(0,0){\strut \emph{system 2}}}
\put(300,32){\makebox(0,0){\strut \emph{system 3}}}
\put(40,60){\makebox(0,0){\strut declarative proofs}}
\put(170,60){\makebox(0,0){\strut declarative proofs}}
\put(300,60){\makebox(0,0){\strut declarative proofs}}
\put(40,130){\makebox(0,0){\strut procedural proofs}}
\put(170,130){\makebox(0,0){\strut procedural proofs}}
\put(300,130){\makebox(0,0){\strut procedural proofs}}
\put(40,122){\vector(0,-1){53}}
\put(170,122){\vector(0,-1){53}}
\put(300,122){\vector(0,-1){53}}
\put(87,60){\vector(1,0){36}}
\put(123,60){\vector(-1,0){36}}
\put(217,60){\vector(1,0){36}}
\put(253,60){\vector(-1,0){36}}
\end{picture}
\end{center}
\xmedskip

\noindent
However, conversions between Mizar style proofs for different systems cannot be completely
automatic.
Even although the \texttt{miz3} language can be
put on top of any system, the semantics of the \emph{statements} (when working
on top of the native library of the systems) will not exactly
match, and the \emph{justification} automation will not match either.
When converting a \texttt{miz3} proof to Mizar, Isar or C-zar,
one can only do an approximate job with the statements,
and there is no proper way to translate all tactics.

However, one gets a very good `starting point' when converting
a declarative proof between systems.
There just will be some justification errors.
This means that a declarative proof converted to a different
system will be like a `formal proof sketch' \cite{wie:04}.

For these reasons our system can also not directly make use
of the Mizar mathematical library MML.
But again, it is not difficult to write a translator that
translates a class of Mizar proofs (the ones that
use statements that map reasonably well to HOL Light versions)
into \texttt{miz3} formal proof sketches.

\section{A larger example: Lagrange's theorem }\label{lagrange}

\noindent
We have tried \texttt{miz3} on a somewhat larger proof: Lagrange's theorem
from group theory.
This states that for any group the order of a subgroup always divides the order of that group.
John Harrison already had written a HOL Light proof of
this theorem, which meant that we could see how the traditional proof style compared
to what was possible in \texttt{miz3}.
We used two different approaches: we wrote a proof following
the proof that is in van der Waerden's book about algebra \cite{wae:71},
and we wrote a \texttt{miz3} proof trying to closely follow John Harrison's proof.

Van der Waerden's proof we formalized both in Mizar and in \texttt{miz3},
and in both cases we first wrote a formal proof sketch \cite{wie:04} of the proof.
The \texttt{miz3} formal proof sketch was:

\xmedskip
\begin{alltt}\small
now let a be A; assume a IN G; let b be A; assume b IN G;
  assume i(a)**b IN H;
  b***H = a**i(a)**b***H; .= a***(i(a)**b***H); thus .= a***H;
end;
!a b. a IN G /{\lam} b IN G /{\lam} ~(a***H = b***H) ==> a***H INTER b***H = \{\}
proof let a be A; assume a IN G; let b be A; assume b IN G;
  now assume ~(a***H INTER b***H = \{\});
    consider g1 g2 such that g1 IN H /{\lam} g2 IN H /{\lam} a**g1 = b**g2;
    g1**i(g2) = i(a)**b;
    i(a)**b IN H;
    thus a***H = b***H;
  end;
qed;
!a. a IN G ==> a IN a***H proof let a be A; assume a IN G; a**e = a; qed;\toolong
\{a***H | a IN G\} PARTITIONS G;
!a b. a IN G /{\lam} b IN G ==> CARD (a***H) = CARD (b***H)
proof let a be A; assume a IN G; let b be A; assume b IN G;
  consider f such that !g. g IN H ==> f(a**g) = b**g;
  bijection f (a***H) (b***H);
qed;
set INDEX = CARD \{a***H | a IN G\};
set N = CARD G; set n = CARD H; set j = INDEX; N = j*n;
thus CARD H divides CARD G;
\end{alltt}
\xmedskip

\noindent
(The notations with the multiple stars is a bit ugly, but we wanted to use the
statement from John Harrison's proof, and there group multiplication
is written as \texttt{**}.
We then used \texttt{***} for multiplication of an element
and a coset.)
Both the Mizar and \texttt{miz3} formalizations were completely
straightforward.
The part of the final formalization that corresponds to the first four
lines of the formal proof sketch became:

\xmedskip
\begin{alltt}\small
now [22]
  let a be A; assume a IN G [23];
  let b be A; assume b IN G [24];
  i(a)**b IN G [25] by 2,23,24;
  assume i(a)**b IN H [26];
  b***H = e**b***H by 2,24;
    .= a**i(a)**b***H by -,2,23;
    .= a**(i(a)**b)***H by -,2,23,24;
    .= a***(i(a)**b***H) by -,9,23,25;
  thus .= a***H by -,17,26;
end;
\end{alltt}
\xmedskip

\noindent
We obtained \texttt{miz3} versions of John Harrison's proof in two
different ways.
We automatically generated this using the technology from Section~\ref{automatic},
but we also wrote one manually by just running the proof step by step, `understanding' what was going on, and then rendering that in \texttt{miz3} syntax.
The first, although a correct declarative proof, was disappointingly large,
because there is a lot of equality reasoning in the proof and the
converter does not optimize that proof pattern yet.
Writing the second again was completely straightforward.

The line counts of the various formalizations that we got were:
\xmedskip
\begin{center}
\begin{tabular}{lr}
\hline
\noalign{\smallskip}
traditional HOL Light by John Harrison & \textbf{214} lines \\
\noalign{\smallskip}
\hline
\noalign{\smallskip}
Mizar formal proof sketch & 25 lines \\
Mizar & 153 lines \\
\texttt{miz3} formal proof sketch & 23 lines \\
\texttt{miz3} & \textbf{183} lines \\
\noalign{\smallskip}
\hline
\noalign{\smallskip}
\texttt{miz3} (converted from John Harrison's proof) & 1,317 lines \\
\texttt{miz3} (manually written following John Harrison's proof) & \textbf{198} lines \\
\noalign{\smallskip}
\hline
\end{tabular}\label{proofcounts}
\end{center}
\xmedskip

\noindent
It is clear that the two hand-written \texttt{miz3} formalizations are of a
similar size to the HOL Light and Mizar formalizations, which shows that
using \texttt{miz3} did not lead to much larger proof texts.

\section{Discussion}\label{discussion}

\subsection{A more generic interface?}

\noindent
The synthesis between proof styles that
we propose in this paper and its accompanying proof language
is very general.
Therefore a natural question is why we did not make the \texttt{miz3}
framework \emph{generic}, like for example the Proof General interface \cite{asp:00}.
That way our work would be useful for users of systems like
Isabelle, Coq and PVS.

The reason we did not do this is that to get a usable
framework we had to do various things that tie deep into the innards
of the system.
Our current implementation really works on the level of the
data structures inside the prover.
And there we do things like timing out tactics, caching justifications,
and interpreting strings as OCaml expressions.

To make a version of our system that is generic, we would need
to work on a much more syntactic level.
However, in that case it probably would not be so easy to do caching of
justifications in a sound way.
We still think this might be possible, but it would be very different from the
architecture that we use now.

\subsection{The reliability of the system}

\noindent
Since we have not changed the HOL Light kernel,
\texttt{miz3} is as reliable as the standard version of HOL Light.
If \texttt{miz3} gives an error message, then \emph{that} might be wrong,
but if a proof is \emph{accepted}, then we know that the
kernel of the system has fully checked the proof object,
and therefore that possible \texttt{miz3} bugs did not matter
(this is called the {de Bruijn criterion} \cite{bar:wie:06}).
This holds even although \texttt{miz3} consists of complicated code,
and even although the system goes outside the basic OCaml
programming model by using a system call to time out tactics and
by invoking the OCaml interpreter on pieces of the proof text.

When a lemma has a completed \texttt{miz3} proof with no errors left,
the proved \texttt{thm} will `appear' in the session for further use.
For example, once we check a correct proof for the example lemma
in Section~\ref{example},
we see magically appear in the HOL Light session:

\xmedskip
\begin{alltt}\small
# val ( ARITHMETIC_SUM ) : thm =
  |- !n. nsum (1..n) ({\lam}i. i) = (n * (n + 1)) DIV 2
\end{alltt}
\xmedskip

\noindent
From then on we can refer to the variable \texttt{ARITHMETIC\char`\_SUM}.
The interface generally suppresses all printing when it is
doing a check
(we do not want a check from \texttt{vi}
to disturb the output of our HOL Light session, where we might be
doing other things),
only when a proof is fully correct the theorem is printed.

To make it possible to get a \texttt{thm} out of a \texttt{miz3} proof,
the justification cache has not only have to hold the
information that a justification was correct, but also a \texttt{thm}
that can be used to redo that justification in a very fast way.

\subsection{Future work}
\label{improvements}

\noindent
The \texttt{miz3} interface is working very well, and is competitive
with the traditional way of using HOL Light.
Still, there are various ways in which it can be improved:

\begin{iteMize}{$\bullet$}
\item
The default prover for \texttt{by} justifications
is not completely satisfactory.
This is orthogonal to the design of the system, but a better
justifier will make the system easier to use.
One would like the justifier to have the following properties:

\begin{iteMize}{$-$}
\item
Any statement that can be proved in HOL Light should be provable in
\texttt{miz3} without explicit tactics.
This does not mean that the \texttt{by} justifier should be able to prove
it all by itself (not even given infinite time and memory), but that it should be possible
to break the proof in sufficiently small steps that \texttt{by}
then can prove each \emph{step} without further procedural help.

\item
Basic HOL Light tactics like \texttt{REWRITE\char`\_TAC}, \texttt{MATCH\char`\_MP\char`\_TAC},
\texttt{ARITH\char`\_{\penalty 100}TAC} and \texttt{MESON\char`\_TAC} should not have to be given
explicitly.
If the tactic in a justification is one of these, and it runs
in the justification in a reasonably short time, then the
default tactic should be able to do the proof too.

\end{iteMize}

\item
Experimenting with goal states that correspond to steps in a \texttt{miz3}
proof could be more ergonomic.
These goal states either can correspond to a \texttt{thesis} at a specific
place in the proof, or to a \texttt{by} justification.
One can experiment already using the \texttt{GOAL\char`\_TAC} tactic (a
custom \texttt{miz3} tactic that
sets the current goal of the HOL Light session to the goalstate it is given as input) but this is
a bit cumbersome.
At the moment Isabelle/Isar is in this respect much more ergonomic than our system.

Currently, when the system flags an error for a certain place in the
proof this gives the user a bit of a helpless feeling
(the same holds for the Mizar system).
If the user understands the problem, then all is fine,
but if the user does \emph{not} understand the error then there should
be something that can be done to investigate more easily than is
possible now.

\item
We might try going beyond the
standard Mizar proof idiom.
For instance, it would be trivial (although not terribly useful) to follow the suggestion from various people to
have an \emph{it now is sufficient to prove this}
proof step.
Such a step sets the \texttt{thesis} and its justification derives the old
\texttt{thesis} from that statement.
It is called `\texttt{suffices} \texttt{to} \texttt{show}' in \cite{har:96} and `\texttt{Suffices}' in \cite{bar:03}.

More interestingly there could be a variant of the \texttt{cases}
construct where the cases are the subgoals produced by a tactic.
That way induction proofs could be more like traditional mathematical proofs,
because then
the \texttt{INDUCT\char`\_TAC} tactic would be \emph{before} the
inductive cases.

\item
It would be interesting to have a more intelligent folding editor on top
of our interface.
It could color the relevant parts of the proof to
clearly show the goal that corresponds to an unjustified step,
and it could intelligently fold and unfold subproofs.

\item
It would be useful to make the communication architecture more flexible.
That way we might have verification not be restricted to one contiguous block,
the system could report errors more precisely,
and several instances of
the system could be run on one computer at the same time.

\end{iteMize}

\subsection{A proof interface for the working mathematician}

\noindent
A good proof interface for formal mathematics should satisfy the
requirement that
\emph{easy things should be easy, and hard things should be possible} \cite{bea:98}.
The synthesis proposed in this paper combines the ease of the declarative
proof style with the power of the procedural proof style.
We hope that our approach will turn out to be a piece in the puzzle
of making interactive theorem provers useful for and attractive to the
working mathematician.

\section*{Acknowledgements}

\noindent
Thanks to Henk Barendregt for his vision of a \emph{luxury mathmode} for interactive
theorem provers, which led directly to the approach from this paper.
Thanks to
Andrea Asperti,
Georges Gonthier,
John Harrison,
Robbert Krebbers,
James McKinna,
Randy Pollack,
Dan Synek,
Laurent Th\'ery and
Makarius Wenzel
for many helpful discussions about this paper.
Thanks to anonymous referees for helpful feedback on various versions of this paper.


\end{document}